# Coverage Games for Testing Nondeterministic Systems

Farn Wang[1,2], Jung-Hsuan Wu[1], Sven Schewe[3], and Chung-Hao Huang[2]

[1] Department of Electrical Engineering, National Taiwan University, Taiwan, ROC
[2] Graduate Institute of Electronic Engineering, National Taiwan University, Taiwan, ROC
[3] Department of Computer Science, University of Liverpool, UK

**Abstract.** Modern software systems may exhibit a nondeterministic behavior due to many unpredictable factors. In this work, we propose the *node coverage game*, a two player turn-based game played on a finite game graph, as a formalization of the problem to test such systems. Each node in the graph represents a *functional equivalence class* of the *software under test* (*SUT*). One player, the *tester*, wants to maximize the node coverage, measured by the number of nodes visited when exploring the game graphs, while his opponent, the SUT, wants to minimize it. An optimal test would maximize the cover, and it is an interesting problem to find the maximal number of nodes that the tester can guarantee to visit, irrespective of the responses of the SUT. We show that the decision problem of whether the guarantee is less than a given number is NP-complete. Then we present techniques for testing nondeterministic SUTs with existing test suites for deterministic models. Finally, we report our implementation and experiments.

## 1 Introduction

*Coverage-based* techniques [11, 12] have been widely used in the management of testing projects of large and complex software systems. The idea is to model the *software under test* (*SUT*) as a finite number of *functional equivalence classes* (*FEC*). The number of FECs that a test plan can cover is then used as an indication of completeness of a verification task and quality of the SUT. For white-box testing, typical test criteria include line (statement) coverage, branch coverage, path coverage, dataflow coverage, class coverage, function/method coverage, and state coverage [11, 12]. For black-box testing, popular criteria include input domain coverage, GUI event coverage, etc. According to the empirical study from [7], coverage-based techniques are effective in detecting software faults.

However, the test coverage of an SUT has to be achieved with various assumptions and nondeterministic responses of the SUT [11, 13]. For example, when we observe how a request message is served by a server SUT, we really have no control whether the server SUT will finish serving the request, deny the request, or be unaware of the request, e.g., due to loss of connection. The best that a verification engineer can do is to use various strategies to try to reach as many FECs of the server SUT as possible. We propose to model this problem as a two player finite-state game [13], which we call a *node coverage game* (*NC-game* for short). The first player is the tester (*maximizer*; he for short) and the second is the SUT (*minimizer*; she for short). The two players play on a finite *game graph* with nodes for the FECs, where the nodes are partitioned into the nodes that are owned by the tester and the nodes that are owned by the SUT. The tester and the SUT together move a pebble from node to node according to the transition relation of the game graph. The tester chooses the next node when the pebble is on one of his nodes, while the SUT chooses the successor when the pebble is on one of her nodes. The objective of the tester is to maximize the number of visited (covered) nodes, while the SUT wants to minimize the number of covered nodes.

An interesting question about NC-games is how much coverage the tester can guarantee, no matter how the SUT reacts to the test input. We call this guarantee the *maximal coverage guarantee* (*MCG* for short). The MCG is calculated under the conservative assumption that the SUT is malicious and tries to minimize the coverage. (For example, a server SUT may decline all interaction requests.) This is common in nondeterministic systems, and a test coverage below the MCG certainly implies deficiency in test execution.

Our two main contributions are as follows.
- We show that the MCG decision problem (the problem whether the SUT can resolve the nondeterminism to prevent a cover over a given threshold $c$) is NP-complete. This complexity is lower than those established for related coverage problems [1,9]. (They are PSPACE-complete, cf. Section 2.)
- For a real-world SUT that may exhibit nondeterministic behavior, we propose techniques that combine game concepts and randomness to apply a static test suite developed for deterministic models to test the SUT. We also report our experimental results for these techniques.

In the remainder of the paper, we first review related work in Section 2 and basic concepts of game theory in Section 3. In Section 4, we define NC-games and introduce the MCG decicision problem, which we prove to be is NP-complete in Section 5. We then present techniques of applying static test suites to SUT with nondeterminism in Section 6 and report on experimental results using our implementation in Section 7.

## 2 Related work

A recent work in test coverage games in is the *proposition coverage game* by Chatterjee, Alfaro, and Majumdar [1]. Their game graph is the same as ours except that each game node is labeled with a set of atomic propositions. The goal of the tester (SUT) is to cover as many (respectively few) propositions as possible. They showed that the decision problem of maximal proposition coverage guarantee is PSPACE-complete.

Another related classic problem is the Canadian traveler problem [9] of Papadimitriou and Yannakakis. A problem instance is a two player game. The players are given a partially observable game graph and do not know which edges are connected. Player 1 may try an edge when he is at the source of the edge. Player 2 then decides whether this edge is connected or not. The connectivity of an edge, once decided, cannot be changed by anyone. The answer to a problem instance is whether the ratio of the traveling distance over the optimal static traveling distance is lower than a given number.

The problem is also PSPACE-complete. In contrast, we show that the MCG decision problem is 'only' NP-complete.

In fact, node coverage has been long accepted in the practice of software testing. Thus our result implies that for software testing, coverage analysis can be achieved without incurring PSPACE complexity.

Nachmanson et al. suggested to use random testing to verify nondeterministic SUTs [8]. They argued that, given enough time budget, in general, random testing can lead to good coverage even without prior knowledge of the MCG. In our experiment, we show that, with some game concepts in the test selection, testing nondeterministic SUTs can be made much more efficient.

## 3 Preliminaries

We first introduce some basic notations. Let $\mathbb{N}$ be the set of non-negative integers. We write $[i, j]$ for the set of integers inclusively between $i$ and $j$. Also, $(i, j]$, $[i, j)$, and $(i, j)$ are shorthands of $[i+1, j]$, $[i, j-1]$, and $[i+1, j-1]$, respectively.

Given a finite set $D$, we use $|D|$ to denote the size of $D$. Given two sets $D$ and $D'$, we use $D - D'$ to denote the difference set of $D$ from $D'$, that is, $D - D' \stackrel{\text{def}}{=} \{d \mid d \in D, d \notin D'\}$.

Suppose that we are given an (infinite) sequence $\phi = v_0 v_1 \ldots$ with elements in a set $V$. For every $i \in \mathbb{N}$, we let $\phi(i) = v_i$. We use $V^*$ to denote the set of finite sequences of elements in $V$. Given two sequences $\phi_1$ and $\phi_2$ such that $\phi_1$ is finite, we use $\phi_1 \phi_2$ to denote their concatenation.

Our NC-game is played on a directed graph, conventionally called a *game graph*. Conceptually, a *node* in the graph represents an FEC of the SUT. An *edge* represents a *transition* between FECs. A node can be owned either by the tester (the maximizer, or player 1) or by the SUT (the minimizer, or player 2). At the beginning of the game, there is a pebble in a dedicated initial node of the graph. Depending on who owns the node that contains the pebble, the owner of the node chooses a transition to move the pebble from the current node to the next node. The interleaving of choices by the SUT and the tester extends a *play*—an infinite sequence of nodes in the game graph. The size of the set of nodes that occur in the play is the *coverage* of the play.

The above-mentioned concepts can be formalized as follows.

**Definition 1.** *(Game graph) A game graph $G = \langle V_1, V_2, E, \nu \rangle$ is a weighted finite directed graph with finite node set $V_1 \cup V_2$, edge (transition) set $E \subseteq (V_1 \cup V_2) \times (V_1 \cup V_2)$, and gain function $\nu : (V_1 \cup V_2) \mapsto \mathbb{N}$. We require that $V_1 \cap V_2 = \emptyset$. Nodes in $V_1$ are owned by the tester while those in $V_2$ are owned by the SUT.* ∎

For convenience, from now on, without saying it explicitly, we assume that we are in the context of a given game graph $G = \langle V_1, V_2, E, \nu \rangle$. Moreover, we let $V$ denote $V_1 \cup V_2$. Figure 1 shows an example game graph $G$. Nodes in $V_1$ are drawn as circles. Those in $V_2$ are drawn as squares. The gain of each node is put down under the node name. A continuous interaction between the SUT and the tester together create an infinite play defined in the following.

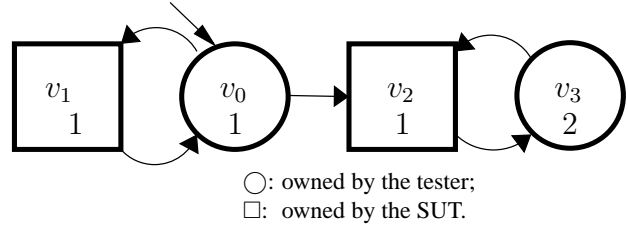

○: owned by the tester;
□: owned by the SUT.

**Fig. 1.** A game graph.

**Definition 2.** *(Plays and play prefixes) A* play $\psi$ *is a function from* $\mathbb{N}$ *to* $V$ *such that, for all* $i \geq 0$, $(\psi(i), \psi(i+1)) \in E$. *A* play prefix $\phi$ *of* $\psi$ *is a mapping from an interval* $[0, k]$ *to* $V$ *such that, for all* $i \in [0, k]$, $\phi(i) = \psi(i)$. ∎

The following notations are for the convenience of presentation. Given a play prefix $\phi : [0, k] \mapsto V$, the *length* of $\phi$, denoted $|\phi|$, is $k + 1$. If $\phi$ is of infinite length, $|\phi| = \infty$. Given two integers $j$ and $k$ in $[0, |\phi|)$ with $j \leq k$, we use $\phi[j, k]$ to denote the play prefix $\phi(j)\phi(j+1)\ldots\phi(k)$. We use $\mathsf{last}(\phi) \stackrel{\text{def}}{=} \phi(|\phi|-1)$ to denote the last node in $\phi$ if the length of $\phi$ is finite ($|\phi| \neq \infty$).

Given a play (or play prefix) $\psi$, we use $[\![\psi]\!]$ to denote the domain of $\psi$, that is, $[\![\psi]\!] = \{\psi(k) \mid k \in [0, |\psi|)\}$. Also, by abuse of notation, $\nu(\psi)$ denotes the coverage gain of $\psi$, i.e., $\nu(\psi) \stackrel{\text{def}}{=} \sum_{v \in [\![\psi]\!]} \nu(v)$.

Without mentioning it explicitly, we assume that a play has infinite length. A play $\psi$ with $\psi(0) = v$ is called a $v$-*play*. In choosing transitions at a node owned by a player, the player may look up the play prefix that leads to the current node, investigate what decisions the opponent has made along the prefix, and select the next node s/he moves to. Such decision-making by a player can be formalized as follows.

**Definition 3.** *(Strategy) A* strategy *is a function from play prefixes to a successor node. Formally, a strategy $\sigma$ is a function from $V^*$ to $V$ such that for every $\phi \in V^*$, $(\mathsf{last}(\phi), \sigma(\phi)) \in E$.*

*A strategy $\sigma$ is* memoriless *(positional) if the choice of $\sigma$ only relies on the current node of the pebble, that is, for every two play prefixes $\phi$ and $\phi'$, $\mathsf{last}(\phi) = \mathsf{last}(\phi')$ implies $\sigma(\phi) = \sigma(\phi')$. If $\sigma$ is not memoriless, it is called* memoriful. ∎

Given regular expressions [6] $\epsilon_1, \ldots, \epsilon_n$ with alphabet $V$ and nodes $v_1, \ldots, v_n \in V$, we may use $[\epsilon_1 \mapsto v_1, \ldots, \epsilon_n \mapsto v_n]$ to (partially) specify a strategy. Supposedly, $\epsilon_1, \ldots, \epsilon_n$ should be disjoint from one another. For a strategy $\sigma$, a rule like $\epsilon_i \mapsto v_i$ means that, for every play prefix $\phi \in \epsilon_i$, $\sigma(\phi) = v_i$. For example, in Figure 1, a memoriless strategy of the tester can be specified with $[V^* v_0 \mapsto v_1, V^* v_3 \mapsto v_2]$. A memoriful strategy of the tester can be specified with $[v_0 \mapsto v_1, v_0 V^* v_0 \mapsto v_2, V^* v_3 \mapsto v_2]$.

Note that, in Definition 3, we do not distinguish between the strategies of the two players. As a player can only influence the decisions made on his or her nodes, we call a play $\phi$ $\sigma$-*conform* for a *tester* strategy $\sigma$ if, for all $i \in \mathbb{N}$, $\phi(i) \in V_1$ implies $\phi(i+1) = \sigma(\phi)$. Likewise, we call it $\sigma$-*conform* for an SUT strategy $\sigma$ if, for all $i \in \mathbb{N}$, $\phi(i) \in V_2$ implies $\phi(i+1) = \sigma(\phi)$.

In the remainder of the paper, we denote the set of all strategies by $\Sigma$. Together with an initial node $r$, two strategies $\sigma_1, \sigma_2 \in \Sigma$ of the *tester* and the SUT, respectively, define a unique play, which is conform to both. We denote this play by $\mathsf{play}(r, \sigma_1, \sigma_2)$.

**Definition 4.** *(Traps) For $p \in \{1, 2\}$, a $p$-trap is a subset $V' \subseteq V$ that player $3 - p$ has a strategy to keep all plays from leaving $V'$. Formally, we require that:*
- *For every $v \in V' \cap V_p$ and every $(v, v') \in E$, $v' \in V'$.*
- *For every $v \in V' - V_p$, there exists a $(v, v') \in E$ with $v' \in V'$.*

*For convenience, in this work, a 1-trap is called a tester trap while a 2-trap is called an SUT trap.* ∎

## 4 Node coverage game (NC-game)

An NC-game is defined with $G$ and an initial node.

**Definition 5.** *(Node coverage game, NC-game) A node coverage game $\langle G, r \rangle$ is a pair of game graph $G$ and an initial node $r \in V$. In the game, the tester (player 1) tries to cover as many nodes as possible in plays while the SUT (player 2) tries to cover as few nodes as possible in plays.* ∎

For convenience, from now on, unless explicitly stated, we assume that we are in the context of an NC-game $\langle G, r \rangle$. The *maximal coverage guarantee* (*MCG*) from $r$ of $G$, denoted $\mathsf{mcg}(r)$, is $\max_{\sigma_1 \in \Sigma} \min_{\sigma_2 \in \Sigma} \nu(\mathsf{play}(r, \sigma_1, \sigma_2))$. Intuitively, this is the maximal coverage gain from $r$ that the tester can guarantee no matter how the SUT may respond.

A strategy $\sigma_1$ of the tester is *optimal* if it can be used by the tester to achieve at least $\mathsf{mcg}(r)$ coverage no matter how the SUT may respond in the game. Formally, $\sigma_1$ is optimal for the tester if and only if $\min_{\sigma_2 \in \Sigma} \nu(\mathsf{play}(r, \sigma_1, \sigma_2)) = \mathsf{mcg}(r)$ holds. Symmetrically, a strategy $\sigma_2$ of the SUT is optimal if and only if $\max_{\sigma_1 \in \Sigma} \nu(\mathsf{play}(r, \sigma_1, \sigma_2)) = \mathsf{mcg}(r)$ holds.

The complexity of a computation problem in computer sciences is usually studied in the framework of decision problem. Our *MCG decision problem* is defined as follows.

**Definition 6.** *(MCG decision problem) Given a $c \in \mathbb{N}$, the MCG decision problem asks whether $\mathsf{mcg}(r) \leq c$.* ∎

## 5 Complexity of the MCG decision problem

This section contains our main theoretical result: we establish the NP-completeness for the MCG decision problem.

**Theorem 1.** *MCG decision problem is NP-complete for both constant and general $\nu$.*

Due to space restrictions, the details of the proofs are moved to the appendix, but we give the main intuition of the proofs below.

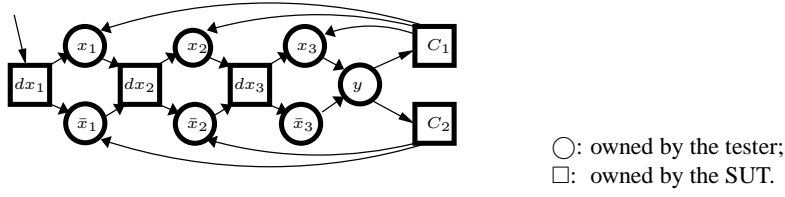

○: owned by the tester;
□: owned by the SUT.

**Fig. 2.** A game graph for $(x_1 \vee x_2 \vee x_3) \wedge (\bar{x}_1 \vee \bar{x}_2)$

### 5.1 NP-hardness

The hardness is easy to establih by a standard reduction from a SAT problem: we reduce from the satisfiability problem of Boolean formulas in conjunctive normal form (CNF) [3] to the MCG decision problem. As outlined in Figure 2, we translate a Boolean formula $\eta$ in CNF with $n$ atomic propositions $x_1, x_2, \ldots, x_n$ and $m$ clauses $C_1, C_2, \ldots, C_m$ into an NC-game $\langle G, dx_1 \rangle$ as follows.

- We have $m + 3n + 1$ nodes, nodes $x_i$, $\bar{x}_i$, and $dx_i$ for $i = 1, \ldots, n$, a node $y$, and a node $C_j$ for $j = 1 \ldots, m$.
- From the nodes $dx_1$, the SUT can choose to go to $x_i$ or $\bar{x}_i$, respectively.
- From the nodes $x_i$ and $\bar{x}_i$, $dx_{i+1}$ is the only successor for $i < n$, and $y$ is the only successor of $x_n$ and $\bar{x}_n$.
- From $y$, the *tester* can choose to go to $C_1, \ldots, C_m$.
- For each clause $C_j$ with $1 \leq j \leq m$ and literal $l$ (some $x_i$ or $\bar{x}_i$) in $C_j$, the SUT can go to $l$.

The size of $G$ is $m + 3n + 1$ and the reduction can be done in polynomial time. If the formula is satisfiable, the SUT can force a cover $\leq m + 2n + 1$: intuitively, she can guess a satisfying assignment and restrict the cover to either the node $x_i$ or the node $\bar{x}_i$ for $i = 1, \ldots, n$ (depending on the assignment), $dx_i$ for $i = 1, \ldots, n$, $y$, and $C_j$ for $i = 1, \ldots, m$. If there is no satisfying assignment, she still has to cover $dx_i$ for $i = 1, \ldots, n$ and either the node $x_i$ or the node $\bar{x}_i$ for $i = 1, \ldots, n$ in the first $2n$ steps, and she cannot prevent coverage of all $C_j$ for $i = 1, \ldots, m$. Moreover, when read as an assignment her choice of nodes $x_i$ or the node $\bar{x}_i$ for $i = 1, \ldots, n$ must violate some disjunction $C_j$. When the *tester* forces the SUT to cover $C_j$, she therefore has to cover one further node $x_i$ or $\bar{x}_i$.

### 5.2 Inclusion in NP

To show inclusion in NP, we show that the SUT has an optimal strategy that can be described in polynomial space and checked in polynomial time. This may look slightly unusual for the testing community, as a strategy would rather be expected for the tester. But a consequence of our results is that the co-problem of determining if a tester has a strategy of a certain quality is CoNP-complete, and we cannot hope to generally have an optimal tester strategy with a similarly simple description.

We start by outlining how an optimal strategy of the SUT can be described, and then give an intuition on why this is the case.

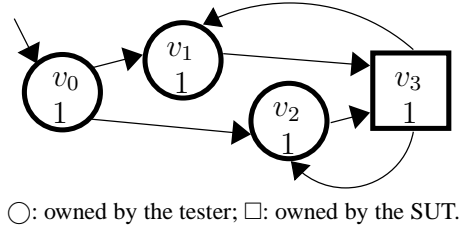

◯: owned by the tester; ▢: owned by the SUT.

**Fig. 3.** Game graphs with memoriful strategies

The key ingredient of our optimal strategy of the SUT is the following. From each node $v$, the SUT would offer the tester a set $P_v$ of nodes that the tester can cover. This set is not necessarily a trap since the tester may be in a position to leave it. With the exception of singleton sets $P_v$, the SUT, however, must have a successor node in $P_v$ for each of her nodes in $P_v$.

A sufficient description of an SUT optimal strategy is to provide, for each state $v$, such a set $P_v$ and the gain $c_v$ the tester can at most obtain against the SUT when starting in $v$. For example, in Figure 3, the SUT must check whether the tester has chosen $v_1$ or $v_2$ to decide whether to transit to $v_1$ or $v_2$ from $v_3$ to contain the coverage at 3 instead of 4. Here we can make the following optimal choice:

- $c_{v_0} = 3$ and $P_{v_0} = \{v_0\}$,
- $c_{v_1} = 2$ and $P_{v_1} = \{v_1, v_3\}$,
- $c_{v_2} = 2$ and $P_{v_2} = \{v_2, v_3\}$, and
- $c_{v_3} = 2$ and $P_{v_3} = \{v_2, v_3\}$.

$c_{v_0} = 3$ reflects the claim that the SUT can restrict the coverage from $v_0$ to three nodes. In order to do so, she first allows the *tester* to cover $P_{v_0} = \{v_0\}$. $P_{v_0}$ is a claim that boils down to "the *tester* is allowed to gain as much as he can in $\{v_0\}$ until he decides to leave $P_{v_0}$ and will not visit $P_{v_0}$ again." Calculating $c_{v_0}$, we find that it is $|P_{v_0}| + \max\{c_{v_1}, c_{v_2}\} = 3$.

From $v_0$, the *tester* can choose to either go to $v_1$ or to $v_2$. This choice determines the next pseudo trap for the *tester* the system moves to, $P_{v_1} = \{v_1, v_3\}$ or $P_{v_2} = \{v_2, v_3\}$, both of which are actually tester traps. The value of $P_{v_3}$ actually does not matter since the choice at $v_3$ is already determined at either $v_1$ or $v_2$.

Note that Figure 3 also serves as an example, where the SUT needs memory for her optimal decisions. This need for memory is reflected in the strategy above: the choice of the SUT is *not* made when $v_3$ is reached, but when $v_0$ is left, either to $v_1$ or to $v_2$. At this point, the memory is taken into account.

It is easy to see that the consistency of such a guess can be checked in PTIME. Together, the hardness and the inclusion argument provide a proof of Theorem 1. The details of the construction is given in Appendix A.

The proof that an optimal strategy can be described in such a simple way uses the existence of *log-consistent* strategies as an intermediate lemma. Log-consistency is a weaker requirement: it allows the selection of a successor of an SUT node $v$ to depend on the history, but if $v$ itself had occurred before, the same successor needs to be chosen as before. Given her objective to minimize the cover, the existence of optimal log-consistent strategies for the SUT is not surprising.

Once we restrict our focus on log-consistent strategies, one can view a run as the building of a directed graph. Let us assume we start in a node $v$ and the SUT follows a log-consistent strategy. Then there is a (not necessarily unique) maximal SCC component that contains $v$ that can be constructed by the *tester*. (This might be the trivial component that contains only $v$ if the *tester* cannot return to $v$ at all.)

The $P_v$ can be chosen to reflect such an SCC component. It is a bit rough in that it only reflects the nodes occurring in this SCC component, but this is enough for the purpose of offering to cover it. Note that, with such an SCC component, the tester could do exactly this: cover it completely and then, leave it in an own node of his choice, unless it is a *tester*-trap or a trivial SCC component that consists only of a single SUT node.

To close the proof, we have shown that a set $P_{v'}$ can be chosen *after* leaving a previously assigned set $P_v$ to a node $v' \notin P_v$ can be selected independent of the history. The details of all these constructions are given in the appendix.

### 5.3 Taking SUT restart into consideration

From the perspective of a tester, one might argue that it should be possible to re-start tests. A simple reduction from the MCG decision problem with restart to the MCG problem without is also provided in Appendix A. As the hardness argument is not affected, we obtain a similar theorem.

**Theorem 2.** *MCG decision problem with re-start is NP complete for both, constant and general $\nu$.*

## 6 Repetitive test selection with game concepts

There are many academic and commercial tools for generating test suites for deterministic models of SUTs. We here present techniques to apply such test suites to nondeterministic SUTs. Conceptually, a *test suite* is a set of finite play prefixes (also called *test cases*) of a game graph. Such test suites can be obtained by viewing the control-flow graph or state-transition diagram of the SUT as a game graph with only one player. For deterministic SUTs, executing each play prefix in the test suite once is enough to get the coverage of the test suite. But for nondeterministic SUTs, this is not the case, since an SUT node in the play prefix may choose a successor different from the one in the play prefix. In traditional settings of software testing, such a diversion from the play prefix is usually interpreted as inconclusive test verdicts or test failure. Thus, straightforward application of such test suites to nondeterministic SUT models may yield low node coverage.

Our idea is to execute those play prefixes, which has been diverted, a few more times in the hope that, when diversion does not always happen, we may cover more nodes. We formalize the idea with the general framework in Algorithm 1. The framework allows us to leverage existing techniques for testing deterministic systems.

**Algorithm 1** Repetitive execution of a test suite.

NT-plan$(T, b)$ // $T$ is a test suite, $b$ is the initial budget, $[\![C]\!] \stackrel{\text{def}}{=} \bigcup_{\phi \in C}[\![\phi]\!]$
1: Let $C = \emptyset$.
2: **while** $b > 0$ and $[\![C]\!] \neq [\![T]\!]$ **do**
3:     Pick a $\phi \in T$ with the greatest *pgain*$(\phi, C, T)$. Update $C$ to $C \cup \{\phi\}$.
4:     Execute $\phi$ and decrement the execution cost from $b$.
5: **end while**

---

Then we can experiment with various testing strategies with different implementations of the function pgain(). Given a test case set $C$, we let $[\![C]\!] \stackrel{\text{def}}{=} \bigcup_{\phi \in C}[\![\phi]\!]$. We also let $\#C$ be the number of distinct nodes in C, i.e, $|\bigcup_{\phi \in C}[\![\phi]\!]|$. In this work, we experiment with the following four strategies. For convenience of discussion, we denote each strategy with a name.

- **s1.5**: the *deterministic iterative strategy*: Simply execute all $\phi$ in T once. This is the deterministic scheme. Specifically, we let *pgain*$(\phi, C, T) \stackrel{\text{def}}{=} 1$ if $\phi$ has not yet been executed; and *pgain*$(\phi, C, T) \stackrel{\text{def}}{=} 0$ otherwise.
- **s2**: the *random strategy*: This is to randomly pick test cases that may still lead to more coverage. Specifically, *pgain*$(\phi, C, T) \stackrel{\text{def}}{=} 1$ if $[\![\phi]\!] \nsubseteq [\![C]\!]$; and *pgain*$(\phi, C, T) \stackrel{\text{def}}{=} 0$ otherwise.
- **s3**: the *random strategy favoring coverage*: This is to randomly pick test cases for more coverage. Formally, *pgain*$(\phi, C, T) \stackrel{\text{def}}{=}$ a random real in $[0, |[\![\phi]\!] - [\![C]\!]|]$.
- **s4**: the *random strategy favoring controlled coverage*: This is to randomly pick test cases that may lead to more coverage with less interference from the SUT. Given a test case $\phi$, we let $\alpha(\phi)$ be the number of positions of SUT nodes in $\phi$, i.e., $\alpha(\phi) \stackrel{\text{def}}{=} |\{i \mid i \in [0, |\phi|), \phi(i) \in V_2\}|$.
  
  *pgain*$(\phi, C, T) \stackrel{\text{def}}{=}$ a random real in $[0, |[\![\phi]\!] - [\![C]\!]|/\alpha(\phi)]$

## 7 Experiments

We have implemented the test plan algorithms introduced in Section 6. We then carried out two experiments to compare their performance with the random walk algorithm [8] and the Ammann and Offutt's static test plan generator[4] with the node coverage criterion. For the first experiments, we have used four benchmarks from real-world projects. The second experiment uses randomly generated game graphs as benchmarks.

### 7.1 Benchmarks from real-world projects

We used the following four benchmarks from the web in our experiment. The first three benchmarks are parameterized with $m$ as a parameter for the concurrency sizes. With parameterized benchmarks, we can collect performance data to see how our techniques scale to the concurrency sizes.

---
[4] http://cs.gmu.edu/~offutt/softwaretest/

- *Inventory system* [2]: The system consists of a server and several clients. The server maintains the inventory of items. The clients help users to retrieve information pertaining to stored items from the server. In an instance of the benchmark, we have $m$ clients and one server process.
- *Chat system* [4]: The system allows clients to enter chat sessions. Clients that have entered a session may post messages. Each posted message is forwarded to all other clients in the same session. Although the postings are delivered in order, the chat system requires only local consistency. In an instance of the benchmark, we have $m$ clients and up to $m-1$ pending messages.
- *Shared multisets* [4]: The system allows simultaneous accesses by several threads and supports insertions, deletions, and queries on multisets. The number of threads sharing the multisets is $m$.
- *Safety Injection System* [5]: This is a simplified version of a control system for safety injection that monitors water pressure and injects coolant into the reactor core when the pressure falls below some threshold.

For the inventory system and the safety injection system benchmarks, we manually constructed the extended finite-state machines of the benchmarks and then used **REDLIB** [10], a model-checking/simulation-checking library, to construct the game graphs. For the chat system and the shared multisets benchmarks, the game graphs are constructed with *Spec Explorer* [4]. The sizes of the game graphs are given in Table 1.

### 7.2 Experiment design

We compare six algorithms in our experiment. To be fair, we set a test budget $b$ dollars for all the algorithms. Visiting a node in a test execution consumes one dollar from our budget. Resetting the test execution, i.e., starting a new test case, happens when the SUT does not respond according to the scenarios prescribed in the current test case and costs $d$ dollars. In this experiment, we let $d = 10$ and examine how the algorithms perform with different budget values.

- **rdm** (for random walk): This is the random walk algorithm by [8] that dynamically and randomly explores the game graphs. The random walks stop when the budget runs out.
- **GMU** (for George-Mason Univ., the school of Ammann and Offutt): This is the test plan generator with node coverage criterion by Ammann and Offutt's software testing tool. The generator accepts a state-transition graph of the SUT and assumes that the SUT is deterministic. The generator then outputs a set of paths from the initial nodes of the graph as the test suite. In our experiment, this algorithm will execute each path in the test suite once. This is very typical of test suite execution for software projects and usually does not consume the whole budget.
- Strategy s1.5, s2, s3, and s4 are respectively executed until the budget runs out.

Also, the SUT makes unbiased random choices of the successor nodes. When a random choice does not match the current test case, a reset operation is incurred.

Since randomness is used in the SUT decision and some algorithms, we carried out 100 runs for each benchmark configuration and algorithm and report the average data.

**Table 1.** Comparison with the GMU graph coverage tool against industrial benchmarks with random moves

| benchmark | $m$ | nodes | budget | GMU | s1.5 | s2 | s3 | s4 | rdm |
|---|---|---|---|---|---|---|---|---|---|
| inventory system with 1 server and $m$ clients | 2 | 20 | 100 | 64.69±0.48% | 72.08±1.81% | 87.19±1.58% | 85.63±1.57% | 84.69±1.41% | 39.26±2.45% |
| | | | 200 | 89.89±0.16% | 85.85±1.59% | 99.04±0.55% | 99.47±0.42% | 99.47±0.42% | 49.85±2.21% |
| | | | 300 | 95.21±0.74% | 90.85±1.37% | 100±0% | 100±0% | 100±0% | 56.67±2.14% |
| | 3 | 80 | 200 | 27.40±0.08% | 47.76±0.46% | 63.57±0.08% | 61.56±0.20% | 58.83±0.31% | 19.96±0.58% |
| | | | 400 | 49.91±0.02% | 66.06±0.35% | 81.04±0.16% | 79.12±0.16% | 78.46±0.20% | 27.01±0.62% |
| | | | 600 | 63.72±0.02% | 77.29±0.40% | 97.93±0.20% | 96.30±0.24% | 95.93±0.24% | 30.76±0.66% |
| | 4 | 308 | 1000 | 27.91±0.01% | 48.15±0.10% | 61.32±0.02% | 59.62±0.04% | 59.33±0.05% | 14.78±0.15% |
| | | | 2000 | 44.79±0.01% | 66.39±0.08% | 81.35±0.04% | 79.57±0.04% | 79.43±0.06% | 21.63±0.16% |
| | | | 3000 | 60.06±0% | 76.98±0.08% | 99.60±0.03% | 98.67±0.05% | 98.36±0.05% | 27.02±0.20% |
| chat system with $m$ clients and $m-1$ pending messages | 4 | 65 | 1000 | 53.10±0.69% | 58.18±0.99% | 81.05±0.89% | 75.33±0.90% | 68.15±0.93% | 48.33±0.86% |
| | | | 1500 | 70.69±0.73% | 70.33±1.01% | 92.41±0.69% | 89.26±0.74% | 82.87±0.98% | 59.46±0.79% |
| | | | 2000 | 72.44±0.63% | 78.77±0.78% | 98.85±0.33% | 97±0.56% | 93.85±0.73% | 65.84±0.79% |
| | 5 | 161 | 6000 | 51.81±0.30% | 55.88±0.31% | 73±0.49% | 67.34±0.45% | 62.80±0.33% | 60.83±0.36% |
| | | | 8000 | 52.80±0.30% | 62.91±0.31% | 83.95±0.50% | 79.13±0.46% | 71.89±0.43% | 69.53±0.43% |
| | | | 10000 | 52.05±0.31% | 66.46±0.33% | 91.65±0.26% | 86.61±0.54% | 78.02±0.40% | 74.79±0.35% |
| | 6 | 385 | 12000 | 29.91±0.11% | 32.27±0.12% | 37.57±0.17% | 35.76±0.11% | 34.20±0.11% | 48.35±0.16% |
| | | | 16000 | 36.22±0.11% | 36.33±0.13% | 43.63±0.20% | 41.93±0.16% | 38.88±0.11% | 55.98±0.14% |
| | | | 20000 | 36.96±0.13% | 39.78±0.13% | 49.64±0.16% | 46.58±0.17% | 43.44±0.13% | 61.48±0.09% |
| shared multisets accessed by $m$ threads | 2 | 19 | 200 | 62.82±1.94% | 67.3±2.41% | 84.77±2.58% | 76.82±3.41% | 77.83±2.85% | 41.31±3.21% |
| | | | 400 | 78.6±2.09% | 80.32±2.46% | 93.94±2.47% | 93.25±2.21% | 91.53±1.9% | 53.11±3.07% |
| | | | 600 | 78.48±2.66% | 89.12±1.55% | 99.06±0.97% | 99.3±0.42% | 99.06±0.65% | 66.28±2.53% |
| | 3 | 43 | 1200 | 73.8±1.17% | 74.42±1.09% | 90.59±1.39% | 88.89±1.12% | 84.91±1.12% | 54.07±1.4% |
| | | | 1600 | 71.56±1.09% | 77.91±1.06% | 94.77±0.78% | 93.13±0.85% | 92.92±0.77% | 59.3±1.17% |
| | | | 2000 | 72.15±0.9% | 81.87±0.85% | 97.57±0.59% | 97.46±0.52% | 94.45±0.9% | 64.46±1.31% |
| | 4 | 91 | 6000 | 65.59±0.39% | 74.75±0.48% | 92.28±0.45% | 89.55±0.5% | 86.18±0.51% | 57.49±0.59% |
| | | | 8000 | 66.01±0.51% | 79.1±0.33% | 95.75±0.4% | 94.41±0.37% | 91.41±0.41% | 62.64±0.55% |
| | | | 10000 | 65.56±0.48% | 82.27±0.3% | 97.5±0.29% | 96.93±0.34% | 95.15±0.39% | 65.73±0.51% |
| safety injection system | - | 99 | 400 | 21.21±0% | 53.07±0.29% | 65.72±0.17% | 59.38±0.25% | 59.07±0.26% | 22.99±0.51% |
| | | | 600 | 28.73±0.05% | 63.55±0.31% | 83.61±0.25% | 75.13±0.23% | 72.17±0.27% | 26.89±0.4% |
| | | | 800 | 35.63±0.05% | 71.53±0.27% | 97.11±0.19% | 88.15±0.23% | 85.88±0.25% | 30.65±0.51% |

### 7.3 For the industrial SUTs

The performance data of the first experiment is in Table 1. We measured the average coverage gains achieved with different budget values. We have also drawn the charts with respect to the table for each benchmark in Figures 4 and 5. In general, our algorithm s2, s3, and s4 usually increase coverage much faster than the others. The only exception is for the chat system with $m = 6$. For this benchmark, the random walk strategy performs best. When we examine the game graph, we found that there are many SUT nodes with many successors. Thus, the test cases from a static test suite can easily run into failure and need reset. In comparison, the random walk algorithm can always accept the choice of the SUT and continue the walk.

### 7.4 For randomly generated game graphs

To avoid our bias in selecting the benchmarks, we also experimented with randomly generated game graphs. We randomly generated 8 game graphs. The performance data is in Table 2. We have also drawn the charts with respect to the table for each benchmark in Figure 6. The experiment data matches our observation in the last experiment, that is,

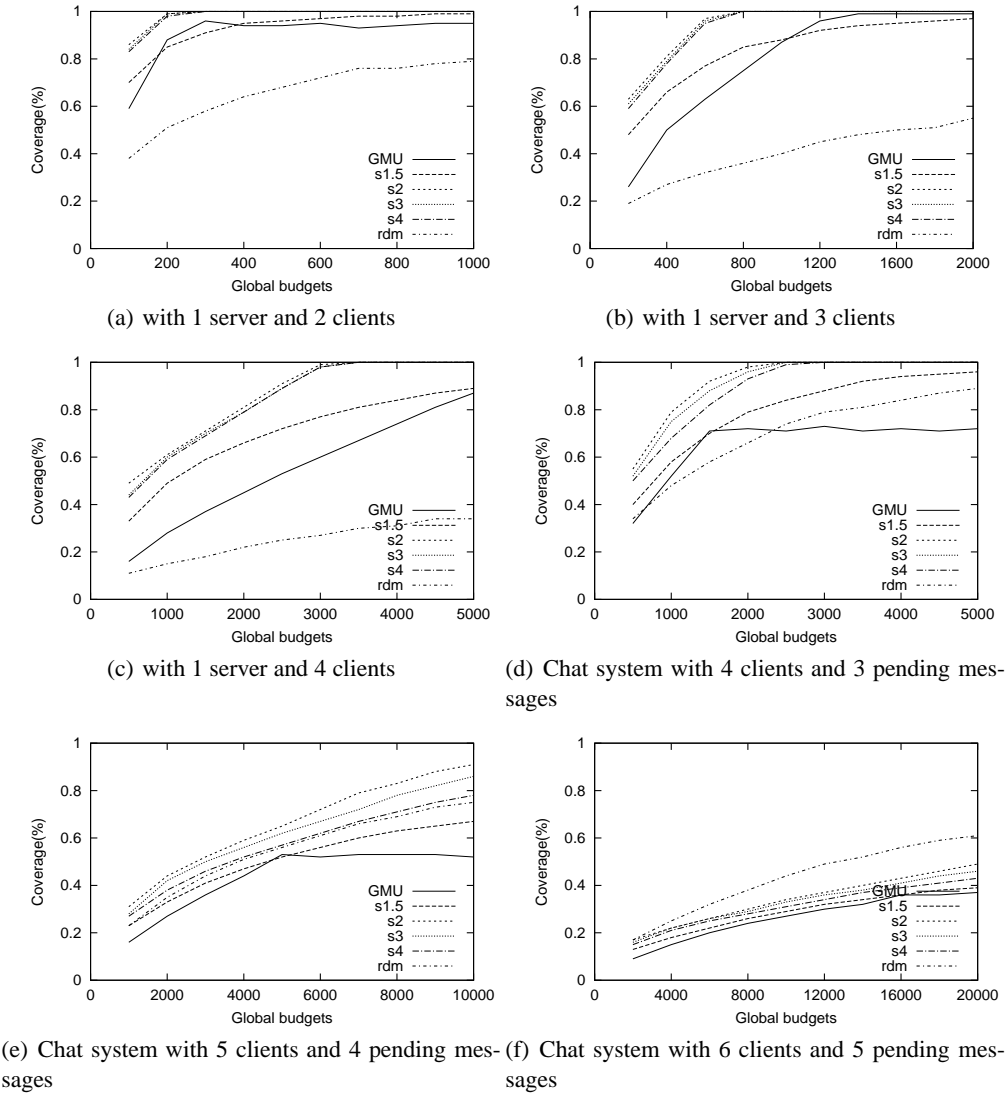

(a) with 1 server and 2 clients

(b) with 1 server and 3 clients

(c) with 1 server and 4 clients

(d) Chat system with 4 clients and 3 pending messages

(e) Chat system with 5 clients and 4 pending messages

(f) Chat system with 6 clients and 5 pending messages

**Fig. 4.** Performance Data of Industrial Benchmarks

algorithm s2, s3, and s4 usually increase coverage much faster than the others. This may imply that testing a nondeterministic system with game concepts could be promising in improving the verification performance.

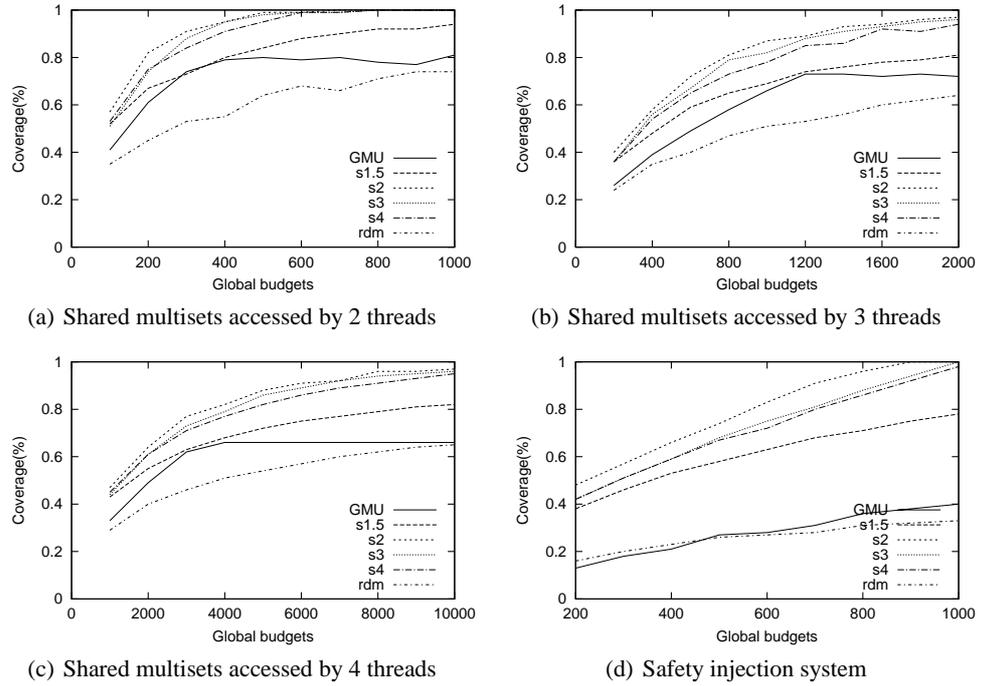

(a) Shared multisets accessed by 2 threads  (b) Shared multisets accessed by 3 threads

(c) Shared multisets accessed by 4 threads  (d) Safety injection system

**Fig. 5.** Performance Data of Industrial Benchmarks (continued)

## 8 Conclusion

We propose to design test plans for covering software products with game concepts. We have established that the MCG decision problem is NP-complete while previous frameworks for game graph coverage [1, 9] are PSPACE-complete. This may imply that our framework of NC-games can bring about more efficiency in various computing aspects of software testing than those of [1, 9]. We have also presented techniques to leverage on existing test plan generators for testing SUTs with nondeterminism. The experimental results are promising and demonstrate the potential of our game-based techniques. We feel that our work not only have laid a concrete foundation for further investigation in this regard, but could also be useful in testing real-world interactive and embedded software systems.

## References

1. K. Chatterjee, L. Alfaro, and R. Majumdar. The complexity of coverage. In *6th Asian Symposium on Programming Languages and Systems (APLAS)*, volume LNCS 5356, pages 91–106. Springer-Verlag, 2008.
2. CONFORMIQ. Company website: http://www.conformiq.com/.
3. S. Cook. The complexity of theorem proving procedures. In *3rd ACM Symposium on Theory of Computing (STOC)*, pages 151–158. Association for Computing Machinery (ACM), 1971.

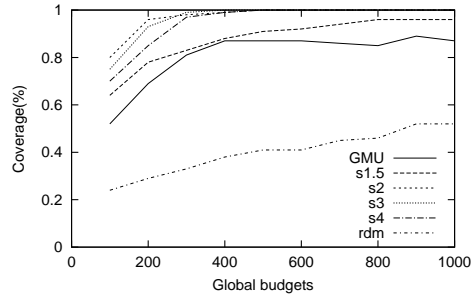
(a) Random game graph 1

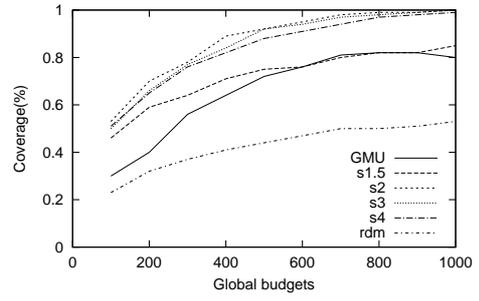
(b) Random game graph 2

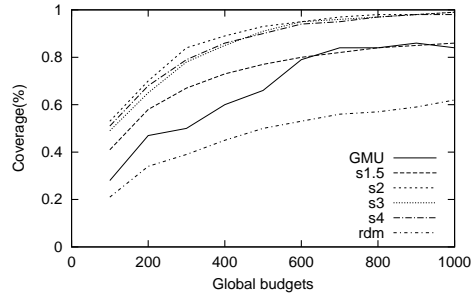
(c) Random game graph 3

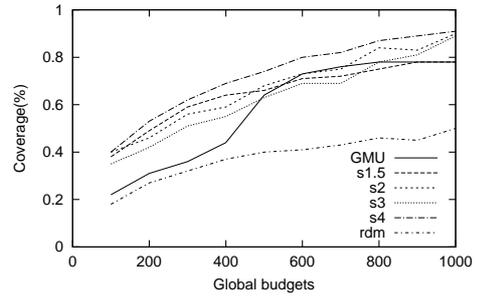
(d) Random game graph 4

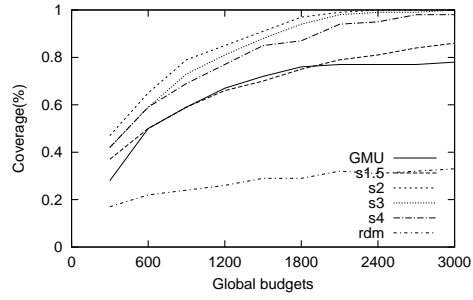
(e) Random game graph 5

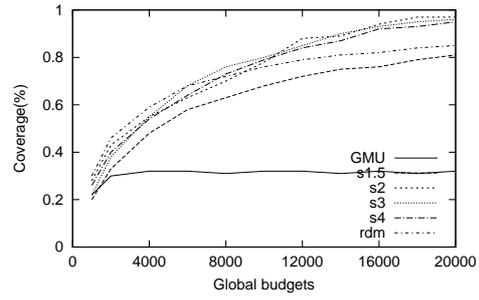
(f) Random game graph 6

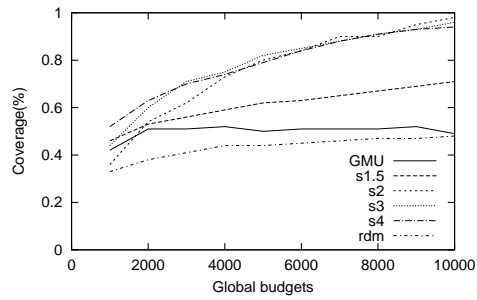
(g) Random game graph 7

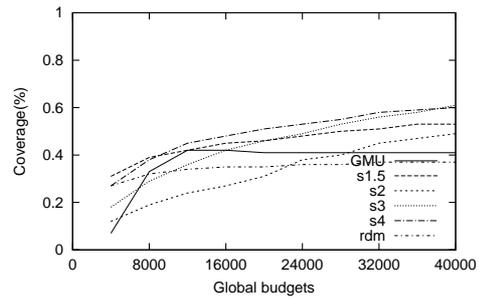
(h) Random game graph 8

**Fig. 6.** Performance Data of Random Game Graphs

**Table 2.** Comparison with the GMU graph coverage tool against random game graphs with random moves

| Game graphs | #nodes | budget | GMU | s1.5 | s2 | s3 | s4 | rdm |
|---|---|---|---|---|---|---|---|---|
| G1 | 19 | 100 | 51.7±0.9% | 64.33±2.52% | 80.47±2.65% | 76.37±2.18% | 70.29±2.33% | 21.53±2.4% |
| | | 200 | 70.33±1.01% | 78.23±1.96% | 97.01±0.95% | 93.42±1.93% | 86.6±2.09% | 29.28±2.33% |
| | | 300 | 80.78±1.56% | 82.99±1.76% | 98.16±1.26% | 99.14±0.59% | 97.06±1.35% | 33.33±3.06% |
| G2 | 30 | 400 | 65.66±0.93% | 72.4±1.12% | 90.08±1.38% | 84.11±1.48% | 82.48±1.54% | 42.22±1.64% |
| | | 600 | 76.51±0.75% | 77.38±1.24% | 94.52±1.65% | 94.13±1.2% | 91.51±1.15% | 45.29±1.1% |
| | | 800 | 81.79±1.01% | 81.37±1.19% | 98.55±0.52% | 98.46±0.56% | 97.44±0.69% | 49.63±1.28% |
| G3 | 40 | 400 | 61.16±0.2% | 73.37±0.87% | 89.48±1.07% | 86.16±1% | 86.51±1.02% | 44.58±1.25% |
| | | 600 | 79.7±0.65% | 80±0.6% | 95.06±0.71% | 95.3±0.62% | 94.35±0.64% | 52.76±1.26% |
| | | 800 | 83.59±0.63% | 83.72±0.59% | 98.46±0.39% | 97.82±0.41% | 97.63±0.57% | 56.48±1.15% |
| G4 | 46 | 600 | 73.86±0.49% | 71.58±1.03% | 78.26±2.77% | 69.41±1.88% | 81.21±1.1% | 39.88±0.79% |
| | | 800 | 78.32±0.87% | 75.03±0.89% | 85.9±2.52% | 78.09±1.86% | 87.74±0.98% | 44.52±0.82% |
| | | 1000 | 77.45±0.75% | 78.94±0.8% | 91.74±1.91% | 89.01±1.53% | 89.94±0.94% | 49.84±0.97% |
| G5 | 75 | 900 | 60.12±0.5% | 59.35±0.58% | 80.81±1.01% | 73.58±0.8% | 69.24±0.67% | 24.58±0.39% |
| | | 1500 | 72±0.7% | 70.95±0.66% | 91.69±1.18% | 90.17±0.87% | 86.39±0.74% | 29.75±0.41% |
| | | 2100 | 77.17±0.5% | 80.37±0.47% | 98.63±0.33% | 98.07±0.32% | 94.23±0.58% | 31.95±0.47% |
| G6 | 95 | 12000 | 31.55±0.58% | 72.83±0.52% | 88.42±1.05% | 85.11±0.76% | 85.06±0.62% | 79.2±0.42% |
| | | 16000 | 30.99±0.58% | 76.98±0.47% | 95.44±0.84% | 93.39±0.57% | 92.2±0.41% | 82.57±0.28% |
| | | 20000 | 31.7±0.53% | 80.63±0.36% | 97.86±0.43% | 96.54±0.38% | 95.97±0.37% | 84.95±0.35% |
| G7 | 100 | 6000 | 50.36±0.41% | 62.86±0.45% | 84.9±0.91% | 86.31±0.64% | 84.26±0.62% | 45.03±0.25% |
| | | 8000 | 51.31±0.55% | 66.95±0.47% | 91.1±0.73% | 92.26±0.53% | 91.46±0.47% | 46.7±0.23% |
| | | 10000 | 49.06±0.37% | 71.29±0.52% | 98.26±0.25% | 97.34±0.38% | 95.8±0.47% | 47.93±0.18% |
| G8 | 297 | 24000 | 41.01±0.15% | 48.53±0.17% | 38.17±0.4% | 49.94±0.25% | 53.46±0.22% | 35.62±0.07% |
| | | 32000 | 41.24±0.16% | 51.59±0.19% | 45.78±0.42% | 56.48±0.24% | 58.54±0.21% | 37.02±0.08% |
| | | 40000 | 41.08±0.18% | 53.73±0.19% | 50.44±0.52% | 62.3±0.23% | 60.71±0.21% | 37.34±0.09% |

4. S. Explorer. Tool webpage: http://research.microsoft.com/specexplorer, January 2005.
5. G. Fraser and F. Wotawa. Test-case generation and coverage analysis for nondeterministic systems using model-checkers. In *2nd International Conference on Software Engineering Advances*, pages 45–50. IEEE Computer Society, 2007.
6. J. Hopcroft and J. Ullman. *Introduction to Automata Theory, Languages, and Computation*. Addison-Wesley, 1979.
7. M. R. Lyu, Z. Huang, S. K. S. Sze, and X. Cai. An empirical study on testing and fault tolerance for software reliability engineering. In *International Symposium on Software Reliability Engineering (ISSRE)*, page 119, Los Alamitos, CA, USA, 2003. IEEE Computer Society.
8. L. Nachmanson, M. Veanes, W. Schulte, N. Tillmann, and W. Grieskamp. Optimal strategies for testing nondeterministic systems. In *ACM SIGSOFT International Symposium on Software Testing and Analysis (ISSTA)*, pages 55–64, New York, NY, USA, 2004. ACM.
9. C. H. Papadimitriou and M. Yannakakis. Shortest paths without a map. *Theoretical Computer Science*, 84:127–150, 1991.
10. F. Wang. Redlib for the formal verification of embedded systems. In *Leveraging Applications of Formal Methods, Second International Symposium, ISoLA 2006, Paphos, Cyprus, 15-19 November 2006*, pages 341–346. IEEE, 2006.
11. E. J. Weyuker. How to judge testing progress. *Journal of Information and Software Technology*, 45(5), 2004.
12. E. J. Weyuker. In defense of coverage criteria. In *Proceedings of the 11th ACM/IEEE International Conf. on Software Engineering(ICSE)*, May, 1989.
13. M. Yannakakis. Testing, optimizaton, and games. In *IEEE LICS*, pages 78–88, 2004.

# APPENDICES

## A  Full version of the Complexity of MCG

We start and end this section with the complexity of the MCG decision problem. It turns out that establishing the lower bound of the MCG decision problem is much simpler than establishing a matching upper bound. We first prove a lemma that comes in handy for establishing the upper bound in Subsection A.2. In the following lemma, we show that the optimal node coverage problem is NP hard.

### A.1  NP-hardness

Note, however, that we show the existence of an SUT strategy; consequently testing the existence of an SUT strategy is CoNP hard.

**Lemma 1.** *The MCG decision problem is NP-hard, even for constant $\nu$.*

*Proof.* We reduce the satisfiability problem of Boolean formulas in conjunctive normal form (CNF) [3] to the MCG decision problem. Suppose we have a CNF formula with atomic propositions $x_1, x_2, \ldots, x_n$ and clauses $C_1, C_2, \ldots, C_m$. We construct a node coverage game $(G, dx_1)$ as follows.
- $V_1 = \{y\} \cup \{\bar{x}_i, x_i \mid 1 \leq i \leq n\}$. $\bar{x}_i$ resp. $x_i$ represent, for each atomic proposition $x_i$, a node that interprets $x_i$ as *false* resp. *true*. $y$ is a special node intuitively used to force the truth valuation of an arbitrary clause (and, ulitimately, of all clauses).
- $V_2 = \{dx_i \mid 1 \leq i \leq n\} \cup \{C_1, \ldots, C_m\}$. Each node $dx_i$ represents a decision node to choose between interpretations $x_i = $ *false* and $x_i = $ *true*. Each node $C_j$ is used for the truth valuation of clause $C_j$.
- $E$ is a minimum set satisfying the following restrictions.
  – For each $1 \leq i \leq n$, $(dx_i, \bar{x}_i) \in E$ and $(dx_i, x_i) \in E$.
  – For each $1 \leq i < n$, $(\bar{x}_i, dx_{i+1}) \in E$ and $(x_i, dx_{i+1}) \in E$.
  – $(\bar{x}_n, y) \in E$ and $(x_n, y) \in E$.
  – For each $1 \leq j \leq m$, $(y, C_j) \in E$.
  – For each clause $C_j$ with $1 \leq j \leq m$ and literal $l$ in $C_j$, $(C_j, l) \in E$.
- $dx_1$ is the starting node of the game.
- For each node $v$ in $V$, $\nu(v) = 1$.

For example, in Figure 2, we have the game graph for a Boolean formula. The size of $G$ is $m + 3n + 1$ and the reduction can be done in polynomial time.

It now suffices to show that a CNF formula $\Psi$ is satisfiable if, and only if, the optimal node coverage with the game graph described is above is $m + 2n + 1$ or smaller (it is precisely $m + 2n + 1$, but this is unimportant for the proof), and unsatisfiable if, and only if, it is strictly bigger.

We prove this claim as follows in two directions. Let us first assume that the minimal coverage is $m + 2n + 1$ (or less). We observe that the minimal coverage must always include all $dx_i$ ($n$ nodes) and $y$ (1 node), because these nodes are always among the first $2n + 1$ visited nodes, irrespective of the chosen strategies, and all $C_j$ ($m$ nodes), because the *tester* can reach each of these nodes from any position of the game. Further, we observe that, within the first $2n$ visited nodes, there is, for all $i \in \{1, \ldots, n\}$, either a visit to node $x_i$ or to node $\overline{x}_i$. These are already $m + 2n + 1$ nodes.

The set of nodes $x_i$ or $\overline{x}_i$ visited within the first $2n$ steps is fixed at the beginning. If the minimal coverage shall be (less than or) equal to $m + 2n + 1$, then the *tester* may not be able to cover any additional node. But as the tester can reach each $C_j$, this means that, for each $C_j$, some successor node (which is either an $x_i$ or an $\overline{x}_i$ node) must have been covered in the first $2n$ steps. According to the construction of the game, the interpretation $I$ from $\{x_1, \ldots, x_n\}$ to $\{true, false\}$ that maps each $x_i$ to *true* if $x_i$ is covered (and to *false* if $\overline{x}_i$ is covered) makes the CNF formula $\Psi$ true.

Vice versa, let us fix such an interpretation $I$. We now consider a strategy of the SUT that moves from $dx_i$ to $x_i$ if $x_i$ is evaluated to *true* by $I$ and to $\overline{x}_i$ if $x_i$ is evaluated to *false* by $I$, and to turn from each $C_i$ to a node $x_j$ (resp. $\overline{x}_j$) such that the respective literal $x_j$ (resp. $\neg x_j$) in $C_i$ is true. Clearly, the coverage is $m + 2n + 1$. ∎

### A.2 Log-Consistent strategies

The main theorem in this appendix is the NP completeness of the MCG decision problem. We then continue with the observation that the complexity of the related problem of testing with re-start (that is, with the capability to re-start the test arbitrarily many times) results in the same complexity.

Before we turn to the NP completeness proof, we will establish that the SUT can play *log-consistent*, that is, it can make the same decision everytime it comes to the same node. Note that this is different to memoryless, as the history at the *first* time it visits $v$ may determine the choice.

For example, in Figure 3, the optimal strategy for the SUT is log-consistent but not memoryless. The SUT must check whether the tester has chosen $v_1$ or $v_2$ to decide whether to transit to $v_1$ or $v_2$ from $v_3$ to contain the coverage at 3 instead of 4. Log-consistent strategies make senses for the SUT since choosing an already-executed transition does not increase the coverage gain for the tester. Log-consistent strategies prove to be sufficient for optimal strategies of the SUT.

The formal definition of log-consistent strategies follows.

**Definition 7.** *(Log-consistent strategies)* A strategy $\sigma \in \Sigma$ of the SUT is log-consistent if, for every $\sigma$-conform play $\phi$ and all $k, l \in \mathbb{N}$, it holds that $\phi(k) = \phi(l) \in V_2$ it holds that $\phi(k+1) = \phi(l+1)$.

For example, with the game graph from Figure 3, a strategy satisfying $[v_0v_1(v_3v_1)^*v_3 \mapsto v_1, v_0v_2(v_3v_2)^*v_3 \mapsto v_2]$ is log-consistent while a strategy satisfying $[v_0\mathcal{V}^*v_3v_1v_3 \mapsto v_2, v_0\mathcal{V}^*v_3v_2v_3 \mapsto v_1]$ is not log-consistent.

**Lemma 2.** *In an NC-game, the SUT has an optimal strategy, which is log-consistent.*

*Proof.* Assume that this is not the case. Let $\sigma_{-1}$ be an optimal strategy in that it guarantees minimal gain $c$ for the SUT, which may or may not be log consistent. We define a sequence of SUT strategies $\sigma_0, \sigma_1, \sigma_2, \sigma_3, \ldots$ as follows.

Let $\phi = v_0v_1v_2 \ldots v_i$ be the prefix of a $\sigma_{i-1}$-conform play. If $v_i \in V_2$, then we choose a $\phi'$ such that

1. $\phi\phi'$ is the prefix of a $\sigma_{i-1}$-conform play,
2. $\phi\phi'$ ends in $v_i$, and
3. among such $\phi\phi'$, the states covered is maximal; that is, irrespective of the strategy of the *tester*, no further state is covered, on any finite sequence that returns to $v_i$.

We use this $\phi'$ to infer $\sigma_i$ from $\sigma_{i-1}$ as follows. We define for every history $\phi\phi_1\phi_2$ with $\mathsf{last}(\phi\phi_1) = v_i$ and $v_i \notin [\![\psi]\!]$ (that is, $\phi_1$ is either empty or ends in $v_i$ and $\phi_2$ does not contain $v_i$), $\sigma_i(\phi\phi_1\phi_2) = \sigma_{i-1}(\phi\phi'\phi_2)$.

We now select the limit strategy $\sigma_\infty = \lim_{n\to\infty} \sigma_i$. $\sigma_\infty$ is well defined, as the reaction on a play prefix of length $i$ is the same as the reaction of $\sigma_i$. The way we update the function clearly ensures log-consistency. We show optimality by induction. As induction basis, $\sigma_{-1}$ is optimal by assumption.

For the induction step, let us assume for contradiction that $\sigma_{i-1}$ is optimal, but $\sigma_i$ is not. Let us assume for contradiction that $\psi = v_0 v_1 \ldots v_i \ldots$ is a $\sigma_i$-conform play with $\nu(\psi) > c$. We now distinguish two cases.

1. $v_i$ occurs infinitely often in $\psi$. But then, by condition (3) of our construction, the nodes covered by $\psi$ are covered by the play prefix $\phi\phi'$ of a $\sigma_{i-1}$-conform play. Consequently, $[\![\psi]\!] \subseteq [\![\phi\phi']\!]$ holds, which implies $\nu(\psi) \leq \nu(\phi\phi')$. As our indcution hypothesis in particular implies $\nu(\phi\phi') \leq c$, this contradicts the assumption $\nu(\psi) > c$.
2. $v_i$ occurs only finitely often, say, it occurs last at position $k \geq i$. Let $\psi' = v_0 v_1 \ldots v_k$ and $\psi = \psi'\psi''$. Then, by the same argument as above, the states covered in $\psi'$ are contained in the states covered by $\phi\phi'$ ($[\![\psi']\!] \subseteq [\![\phi\phi']\!]$). Also, by construction, $\phi\phi'\psi''$ is a $\sigma_{i-1}$-conform play. We thus obtain $[\![\psi'\psi'']\!] \subseteq [\![\phi\phi'\psi'']\!]$, and consequently $\nu(\psi) \leq \nu(\phi\phi'\psi'')$. As our indcution hypothesis provides $\nu(\phi\phi'\psi'') \leq c$, this contradicts the assumption $\nu(\psi) > c$.

After having established that all $\sigma_i$ are optimal, let us assume for contradiction that $\sigma_\infty$ is not, that is, that there is a $\sigma_\infty$-conform play $\phi$ with $\nu(\phi) = c' > c$. But then, some finite initial sequence of $\phi'$ of $\phi$ must have a gain $c' = \nu(\phi')$, and $\phi'$ must be a prefix of a $\sigma_{|\phi'|}$ conform play $\psi$. Consequently $\nu(\psi) \geq \nu(\phi') > c$ holds (contradiction). ∎

### A.3 Inclusion in NP

Having established that it is enough to consider log-consistent strategies, we outline an algorithm that guesses a witness SUT strategy that includes, for each node $v$ of the game,

- the coverage of $c_v$ that the SUT allows the tester to obtain if we start from this node (that is, on a game that is modified only in that the initial node is changed to $v$), and
- a set $P_v \subseteq V$ (for pseudo trap from $v$) of nodes that includes $v$ and satisfies the following constraints:
  - $P_v = \{v\}$ is singleton and $v$ is an SUT node, or
  - for SUT node $u$ in $P_v$ there is a node $w$ in $P_v$ such that $(u, w) \in E$.

For a game that starts in $v$, the SUT would intuitively offer the *tester* to cover $P_v$. The game is then intuitively divided into two phases: the 'covering phase', where $P_v$ may be covered by the *tester*, followed by a phase where the game would not return to $P_v$.

We will argue that the SUT can play optimal by playing from a node $w$ reached after $P_v$ is left as if the game would start in $w$. Not returning to $P_v$ would merely be a property of an optimal SUT strategy that starts in $w$ and not technically required.

A strategy of the SUT is called a *witness strategy* if the SUT can follow the strategy
- to stay in $P_v$ by playing, from a node $u \in P_v$, an edge $(u, u')$ with $u' \in P_v$, and
- to follow the strategy for starting in $w$ as soon as $P_v$ is left to $w$.

The SUT may use this witness strategy to trap plays in $P_v$ until the *tester* decides that he cannot gain more coverage in $P_v$ and had better leave $P_v$ for $w$.

A witness is *consistent* if it satisfies the following side constraints:

- **Case S** (for singleton): if $P_v = \{v\}$, $v$ is an SUT node, $v$ is no sink, and the self-loop $(v, v)$ is not an edge of the game, then we have $c_v = \min\{c_u + 1 \mid (v, u) \in E\}$, and
- **Case T** (for trapping): otherwise, let $c_v = \max\{c_w \mid u \in P_v \cap V_1, (u, w) \in E, w \notin P_v\} + |P_v|$, where the maximum over the empty set is 0 (to account for the case that $P_v$ is a *tester* trap).

Consistency is easy to check.

**Lemma 3.** *Consistency can be checked in polynomial time.* ∎

To follow a consistent witness, the SUT needs a set variable $T$ to record the current set of nodes that the tester is allowed to cover. The following steps describe how a consistent witness strategy can be followed. Initially, $T$ is $\emptyset$. At a node $v$, the SUT first updates $T$ according to the following cases.
- If $v$ is a case S node, $T$ is set to $\emptyset$.
- If $v$ is a case T node, the following cases are further considered.
  - If $T$ is $\emptyset$, then $T$ is set to $P_v$. (This refers to the case that $v$ is the initial node.)
  - If $v \in T$, then we are still in the same set of nodes that the witness strategy has allowed the tester to cover. Thus, we ignore $P_v$ and let $T$ stay unchanged.
  - If $v \notin T$, then we just left the last set of nodes that the witness strategy allowed the tester to cover. Thus, we reset $T$ to $P_v$. Note that $v$ may be a tester node.

Then at a node $v \in V_2$, the SUT makes the following decision.
- If $T = \emptyset$, this implies that $v$ is a case S node and the SUT should pick a successor $u$ with $(v, u) \in E$ and $c_u = \min\{c_w \mid (v, w) \in E\}$.
- If $T \neq \emptyset$, the SUT should choose a $u \in T$ with $(v, u) \in E$.

**Lemma 4.** *If $\nu(v) \geq 1$ for all $v \in V$ then the SUT can, for every consistent witness, guarantee a gain of at most $c_v$ from every node $v \in V$.*

*Proof.* We show this by induction over $c_v$. For the induction basis, this is clearly true if $P_v$ is a *tester* trap, as this implies that $\{c_w \mid u \in P_v \cap V_1, (u, w) \in E, w \notin P_v\}$ is empty, and consequently $c_v = \nu(P_v)$ holds. Note that this is in particular the case for $c_v = 1$.

For the induction step, we can follow the witness strategy to obtain all guarantees smaller than $c_v$ by induction hypothesis. We now distinguish two cases.

1. **Case S** of $v$: Then there must be a successor $u$ (with $(v, u) \in E$) with $c_u = c_v - \nu(v) < c_v$. By induction hypothesis, the SUT can restrict the gain to $c_u$ using the witness strategy, and consequently to $c_v$ from $v$.
2. **Case T** of $v$: Following a witness strategy allows to cover first (some or all) nodes in $P_v$, and then might continue to a successor $w \notin P_v$ of any *tester* node in $P_v$. (Recall that a witness strategy of the SUT will not leave $P_v$ from any SUT node.) We have $c_v > \max\{c_x \mid u \in P_v \cap V_1, (u,x) \in E, x \notin P_v\} \geq c_w$ by the consistency of the witness. Thus, the tester can guarantee a gain of at most $c_w$ after moving to $w$, while clearly at most $\nu(P_v)$ has been gained before. ∎

Note that the correctness argument does not require that $P_v$ is not visited again. It is indeed possible to construct consistent witnesses who do this – not all consistent witnesses are optimal. We now show that an optimal consistent witness exists.

**Theorem 3.** *For a game with $\nu(v) \geq 1$ for all $v \in V$ where an optimal SUT can restrict the gain to $m_v$ when starting in node $v$, there is a consistent witness with $c_v = m_v$ for all nodes $v$.*

*Proof.* We show by induction that, for games with a minimal gain of $m_v$, we can construct a witness with $c_v = m_v$ for all $v \in V$. By the previous lemma, this shows that the SUT can force that, when starting at $v$, the gain is bounded by $m_v$.

For the *induction basis*, this is clearly true if there is a tester trap $P_v$ with $m_v = \nu(P_v)$. For such tester traps, we can simply select $c_v = \nu(P_v) = m_v$. Note that this in particular includes the case $m_v = 1$.

For the *induction step*, we exploit that log-consistent strategies are sufficient for the SUT (Lemma 2). Let us assume that $m_v$ is the correct value, and that the SUT uses a fixed log-consistent optimal strategy. Then, we can look at a run as a producer of edges in a directed graph: we start with the game intersected with the transitions whose source is a *tester* node. That is, we remove exactly the transitions that exit an SUT node.

During the run, every time we pass by an SUT node, we add the selected transition to the di-graph. For a run, we choose the maximal of these graphs such that the added transition belongs to the same SCC as the starting node $v$.

For every run, there is obviously one such graph, and there is obviously a (not necessarily unique) graph among them where this SCC is maximal. We select such a graph $G$ and select $P_v$ to be the set of nodes in the same SCC as $v$. For the SUT nodes in this SCC, we memorize the (due to log-consistency unique) outgoing edge selected by the strategy.

Moreover, we pick, for each successor $w$ of a *tester* node in $P_v$, a history $h_w$ consistent with our log-consistent optimal strategy such that (1) initially $P_v$ is covered completely and (2) then a transition to $w$ is taken. It is obvious that such a history exists, as the play in which $P_v$ is first an SCC can be extended to cover $P_v$ and then to move on to $w$. From $w$, the SUT will play as if it used the old strategy from $v$ and had previously seen $h_w$. We call the inferred strategy of the SUT to continue from $w$ $w$-consistent. Note that a $w$-consistent strategy is (1) log-consistent and (2) guarantees that $P_v$ is not visited again, as $P_v$ is a maximal SCC.

Note that the following strategy provides the same guarantees: while in $P_v$, stay in $P_v$, using the transitions from $G$; once $P_v$ is left to a node $w$, follow the $w$-consistent strategy. It is therefore an optimal log-consistent strategy from $v$. (Optimal because the SUT has the described strategy to achieve the same guarantees as under the optimal log-consistent strategy we started with.)

Note that, by construction, $P_v$ is not reachable from $w$ under the $w$-consistent strategy. (If it was, the SCC was not maximal.) Consequently, we have $c_w < c_v$, and, by *induction hypothesis*, we verify the correct $c_w$ using our witnesses.

Let us assume for contradiction that this is smaller than the result using the $w$-consistent strategy from $w$. Then the strategy from above can be strictly improved to: while in $P_v$, stay in $P_v$, using the transitions from $G$, once $P_v$ is left to $w$, follow an optimal strategy that starts in $w$. This provides a contradiction to the optimality of this strategy.

This leaves the special case where $P_v = \{v\}$ contains a single SUT node without a self-loop, but with some successor. In this case, we have a clearly defined $w$-consistent strategy for the successor $w$ selected by the SUT (and memorize the edge $(v, w)$, together with the guarantee that, in the $w$-consistent strategy, $v$ cannot be visited. This provides us with a strategy for the SUT to restrict the gain from $w$ to $c_v - \nu(v)$. ∎

Note that the restriction to $\nu(v) \geq 1$ is no restriction. For a game with $n$ nodes, we can use the function $\nu'$ with $\nu'(v) = 1 + n \cdot \nu(v)$ instead: a strategy is optimal for $\nu$ if it is optimal for $\nu'$ (thought the converse does not hold).

Putting the lemmata of this section together, we obtain inclusion in NP with this observation. With the matching hardness result from Lemma 1, we get our main theorem.

*Theorem 1.* MCG decision problem is NP complete for both, constant and general $\nu$.
∎

A slight variation of the problem is to allow the *tester* to re-start the game at any time. This variation is natural as one can argue that a test can be re-started.

We first note that the hardness proof is not affected: the SUT can repeatedly guess the same satisfying assignment for the CNF SAT problem.

For the inclusion, we can consider two games: the game with re-set, and a game where every node is doubled into an in- and an out-node, both with the same gain. A transition from $v$ to $w$ becomes a transitions from the out-node of $v$ to the in-node of $w$, and the new initial node is the in-node to the old initial node. The in-node of $v$ has two outgoing transitions: one to the initial node and one to the out-node of $v$. While all in-nodes are *tester* nodes, the out-node of $v$ belongs to the player that owned $v$.

The only other change applied is that, if $v$ is a sink in the game with re-set, then we add a transition from the out-node of $v$ to the new initial node.

It is now easy to show that the maximal gain of the re-start game is exactly half the maximal gain of the new game. We can show this by simulating the games.

*Theorem 2.* MCG decision problem with re-start is NP complete for both, constant and general $\nu$.
∎